# Compact highly efficient 2.1-W continuous-wave mid-IR Fe:ZnSe coherent source pumped by Er:ZBLAN fiber laser

A.V. Pushkin[1], E.A. Migal[1], H. Uehara[2], K. Goya[2], S. Tokita[2], M.P. Frolov[3], Yu.V. Korostelin[3], V.I. Kozlovsky[3], Ya.K. Skasyrsky[3] and F.V. Potemkin[1]

[1] *Faculty of Physics and International Laser Center M.V. Lomonosov Moscow State University, Leninskie Gory, bld.1/62, 119991, Moscow, Russia*
[2] *Institute of Laser Engineering, Osaka University, 2-6 Yamadaoka, Suita, Osaka 565-0871, Japan*
[3] *P.N.Lebedev Physical Institute, Russian Academy of Sciences, Leninsky prosp. 53, 119991, Moscow, Russia*
*Corresponding author: av.pushkin@physics.msu.ru, potemkin@physics.msu.ru*

**We report the compact and robust coherent source operating in mid-IR based on Fe:ZnSe chalcogenide gain medium optically pumped by Er:ZBLAN fiber laser. The output power of 2.1 W with 59% slope efficiency with respect to absorbed pump power at liquid nitrogen cooling is achieved. We show that strong re-absorption at high pump power and iron ion doping concentrations leads to the continuous tuning of central wavelength from 4012 to 4198 nm. Robustness of high power Er:ZBLAN fiber laser combined with prominent spectroscopic properties of Fe:ZnSe media pave the way for the development of reliable tunable CW mid-IR sources for scientific and industrial purposes.**

http://dx.doi.org/10.1364/OL.99.099999

In the last decade, intense efforts have been addressed to the development of near- and mid-IR laser sources based on chalcogenide crystals doped by transition metal ions. Such lasers have found numerous scientific and industrial applications in different areas including molecular spectroscopy, remote sensing, laser radars etc. [1], [2]. Among the whole family of transition metal doped crystals Cr- and Fe- doped ZnSe crystals are the most successful and elaborated ones, lasing in 2 and 4 µm spectral regions, respectively. The rapid development of CW Cr:ZnSe laser was facilitated by the availability of powerful CW pump sources, especially Er- and Tm- fiber lasers. Nowadays output power as high as 140 W at 2.5 µm was achieved from Cr:ZnSe laser [3]. At the same time, the development of iron doped lasers was limited by the lack of powerful and convenient 3-µm pump sources [4]. For operation in gain-switching regime the following solid state and chemical lasers were used: 2.698-µm Er:YAG [5], 2.936-µm Er:YAG [6], [7] 2.85-µm Cr:Yb:Ho:YSGG [8], [9], 2.79-µm Cr:Er:YSGG [10], HF laser [11], Raman shifted Nd:YAG[12]. Continuous-wave generation of Fe:ZnSe was demonstrated only under cryogenic cooling. For the first time, CW operation of Fe:ZnSe gain medium was reported in [13], where Cr:CdSe laser tuned to a wavelength of 2.97 µm was used as a pump source. At this set-up, 160 mW output power at 4 µm was reached. In [14] 840 mW output was achieved using 2.94-µm microchip pump lasers with a total power of 3 W. One more pump source Er:$Y_2O_3$ was introduced in [15] with an output power of 420 mW at 4 um. Nowadays the highest power in continuous-wave regime in Fe:ZnSe was obtained with the use of a cascade scheme based on Cr:ZnSe [16] (described in detail in [3]). This approach resulted in a 9.2 W output power with 41% efficiency. The remarkable laser properties of chalcogenides provide high output power with a wide tuning bandwidth. The special spectral selective design allows for obtaining up to 32 W at 2.94 µm output power for pumping Fe:ZnSe crystal [3]. Such powerful solutions are commercially available and provide quite a high efficiency. Nevertheless, to increase the total optical-to-optical efficiency, powerful CW three-micron laser sources for direct pumping of Fe:ZnSe are more preferable since do not require intermediate gain media (such as Cr:ZnSe or Cr:CdSe). Table 1 presents the main characteristics of reviewed systems.

A new era in the development of three-micron CW laser sources was opened by Er:ZBLAN fiber technology [17]. In continuous wave mode, fiber lasers are more convenient and reliable than solid-state lasers because of the large surface-to-volume ratio. Despite the poor thermal characteristics of the mid-IR fluoride fiber, profound thermal management allowed such sources to come out at a power of 20-30 W [18]. This became possible due to the advanced technology of end-caps, preventing fiber damage caused by high absorption of radiation in atmosphere water vapor [19]. Er:ZBLAN operation in CW mode with output power up to 24 W [18] and Q-switched mode [20] were demonstrated. These lasers are compact and robust, providing high output power and excellent beam quality inherent to fiber lasers.

**Table 1. State-of-the-art on continuous-wave Fe:ZnSe**

| Link | Pump source | Intermediate gain medium | Fe:ZnSe output power | $\eta_{int}$ | $\eta_{pump}$ |
|---|---|---|---|---|---|
| This paper | Er:ZBLAN (2.8 µm, 6.5 W) |  | 2.1 W |  | 34% |
| [13] | TDFL (1.9 µm, 3 W [*]) | Cr:CdSe (0.6W at 2.97 um) | 160 mW | 27% | 5.3% |
| [14] | Two microchip Er:YAG lasers (total 3W at 2.94 um) |  | 840 mW |  | 28% |
| [15] | Er:$Y_2O_3$ (2.74 um, 3.15 W) |  | 340 mW |  | 11% |
| [22] | TDFL (1.9 µm, 15 W) | Cr:ZnSe (5.5 W at 2.94 um) | 1.6 W | 30% | 11% |
| [16] | TDFL (1.9 µm, 110 W) | Cr:ZnSe (32 W at 2.94 um) | 9.2 W | 41% | 13% |

$\eta_{int}$ - efficiency with respect to the pump source
$\eta_{pump}$ - efficiency with respect to the intermediate medium
[*] – data from private conversation

In this paper, we present a mid-IR continuous wave 2.1-W laser based on Fe:ZnSe crystal pumped, for the first time, by a compact 2.8-µm Er:ZBLAN fiber laser.

A schematic diagram of Er:ZBLAN laser is sketched in Fig. 1. A double-clad Er:ZBLAN fiber with a length of 1.8 m (fabricated by FiberLabs, Inc.) is used as the active medium. The specifications of the fiber core are a diameter of 33 µm, an NA of 0.12, and an Er- ions concentration of 6 mol. %. Due to the high level of doping, high output power is achieved in a relatively short fiber, allowing better thermal management. The inner and outer claddings of the fiber have a diameter of 330 µm (octagonal) and 460 µm respectively. For a thermal management of the fiber, a water cooled aluminum radiator was designed. The laser cavity was formed by the concave rear mirror with a focal length of 15 mm and the output facet of the fiber. A fiber-coupled wavelength stabilized laser diode (IPG Photonics) operating at 975 nm is used as the continuous-wave pump source. The collimated pump beam is sent into the inner cladding of the fiber through spherical YAG lens with an 8 mm focal length and AR coatings for both the pump and laser wavelengths (975 nm and 2.8 µm). The dichroic mirror has a high transmittance of 95% at 975 nm and high reflectance of 99.8% at 2.8 µm for incident angles close to 45° and was used to separate the laser output beam. The input-output characteristics are shown in Fig 1. The maximum output power of 6.5 W at a wavelength of 2.80 um was obtained at 35 W pump power with a slope efficiency of 19.9%. An increase in output power is possible by optimizing the input of pump radiation into the fiber cladding or use both-side pumping. Such a powerful laser with foot-print of only 40x60 cm is compact, stable and reliable.

Two Fe:ZnSe single-crystal active elements with $2.5\times10^{18}$ and $3.5\times10^{18}$ cm$^{-3}$ dopant concentrations were used in the experiments. Both crystals were grown from the vapor phase on a single-crystal seed by using the concurrent-doping technology [21]. Crystals grown by this technique possess high structural quality, optical homogeneity and, as a result, low intrinsic losses. The low doped element was 8.5-mm long with AR-coated 10.0x10.0-mm facets. The high doped crystal was uncoated, 8.0-mm thick with 8.0x3.0-mm working facets. $Fe^{2+}$ ions concentration in active elements was determined experimentally from the pump absorption measurements at room temperature. Absorption cross section at λ=2800 nm ($\sigma=0.87\times10^{-18}$ cm$^2$) was taken from [22].

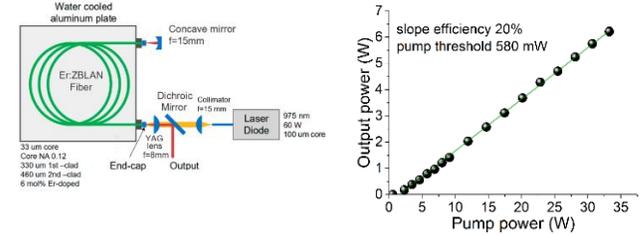

**Fig 1.** A schematic diagram of Er:ZBLAN fiber laser and its laser performance.

The lifetime of the upper laser level in a Fe:ZnSe crystal at room temperature is as short as 370 ns [23] which makes continuous-wave generation sophisticated and demanding high pump power. Therefore, in order to increase the population inversion and obtain laser generation, the crystal was mounted in a liquid nitrogen cooling cryostat. The lifetime at 77 K is about 60 µs [24]. The decrease in the lifetime with temperature raise is caused by phonon-assisted quenching of the luminescence [24].

The experimental setup of Fe:ZnSe laser is depicted in Fig. 2. The design of cooling system makes it possible to mount the entire laser cavity inside cryostat that avoids intracavity losses due to reflection from the input windows, and allows for alignment of the output mirror and adjustment of the cavity length. One window of the cryostat was a plane-parallel plate of the $CaF_2$ plate, while the other one was the cavity output coupler.

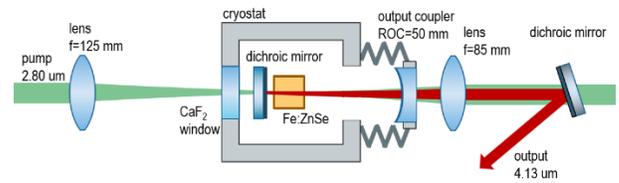

**Fig.2.** Schematic diagram of the experimental setup.

The cavity was formed by a flat dichroic mirror and a curved (ROC=50 mm) output coupler. The dichroic mirror had a high transmittance at the pump wavelength (T=95% at 2800 nm) and high reflectivity at output wavelength (R>99.5% for 3900-4900 nm). For the sake of comparison, two output couplers were used with transmission of about 17% and 34% at 4 – 4.2 µm wavelengths. Both of them possess reflectivity of 18% at the pump wavelength. The geometry of the resonator was close to semi-concentric to provide the smallest mode diameter and the highest gain. The pump radiation was focused by a $CaF_2$ lens with a focal length of 125 mm into 250-µm spot measured at $1/e^2$ intensity level. The Fe:ZnSe crystal was oriented normally to the optical axis of the cavity and mounted at a distance of about 1 mm from the flat mirror. The pump beam was coupled along the optical axis of the cavity through the cryostat window and the dichroic mirror. Because of the high absorption of pump radiation at a wavelength of 2.8 µm in the atmosphere water vapor, only 84% of the Er:ZBLAN output power was delivered to the cryostat input window. Reflection losses of coupling optics and Fe:ZnSe crystal facet were 28.5% summarily. Radiation at a wavelength of 4 µm was separated from the residual pump with a mirror similar to a dichroic cavity mirror.

The dependence of output power on absorbed pump power for both Fe:ZnSe crystals is depicted in Fig. 3. The absorbed pump power was calculated from transmission measurements and took into account reflections from the optical elements. Using 17% output coupler the power of 1.49 W and 1.84 W was achieved for low and high doped crystals respectively. The slope efficiency with respect to the absorbed pump power was almost 50% in both cases. Higher dopant ion concentration allows for absorbing the greater part of pump radiation and thus reaching higher output power. The threshold pump power was about 40 mW in both cases. The 34% output coupler enabled to increase the output power up to 2.13 W and slope efficiency up to 59% using the high doped sample. No thermal roll-off at the high pump power was observed that paves the way for further power scaling. Output beam profile acquired with Spiricon Pyrocam III is shown in the inset of Fig. 3. It represents a good beam quality at TEM$_{00}$ operation.

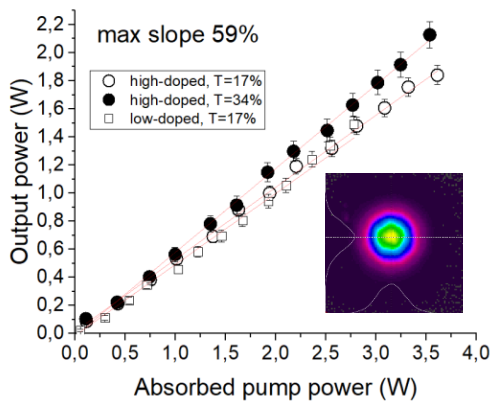

**Fig.3.** Performance of Fe:ZnSe laser based on high doped (N=3.5x10$^{18}$ cm$^{-3}$) and low doped (N=2.5x10$^{18}$ cm$^{-3}$) samples and output beam profile.

Spectral measurements were carried out using a scanning monochromator with a 300 groove/mm grating with a spectral resolution better than 1 nm. Continuous-wave radiation was gated with a mechanical chopper to use sensitive PbSe detectors. The measured spectra have a spiky structure indicating operation on many longitudinal modes of the cavity during the measurement time.

Output spectra for high doped Fe:ZnSe crystal at various output power is shown in Fig. 4. Increase in output power from 200 mW to 1.2 W is accompanied by the red shift of central wavelength from 4149 nm to 4198 nm, which is associated with local heating of the crystal in the pumping region. As a result, a redistribution of the population among the energy levels takes place causing increasing probability of re-absorption. With an increase of temperature, the absorption spectrum broadens mainly to the red region, while the luminescence spectrum expands to the blue wing [24]. The overlap between these spectra leads to strong re-absorption at short wavelengths and, as a consequence, vanishing of blue components from the output spectrum. This phenomenon was also observed in Fe:ZnSe media in [14] and [25].

To evaluate the influence of concentration-dependent re-absorption, spectra of the low doped sample were measured for the same output power. As shown in Fig. 4, low doped crystal demonstrates broader spectrum placed at shorter wavelengths compared to high doped sample. This fact indicates that high doping levels shift spectrum of output radiation to the long-wavelength region irrespective of the absorbed power.

In conclusion, we, for the first time, demonstrate efficient fiber Er:ZBLAN pumping of cryogenically cooled continuous-wave Fe:ZnSe crystal with 2.1-W output power. The slope efficiency of 59% with respect to absorbed pump power was obtained. Measured Fe:ZnSe output spectra indicate a significant influence of re-absorption on generation wavelength. For high doping levels and output powers, spectrum shifts to the red wing which makes possible continuous tuning from 4012 to 4198 nm. The total output power of developed source can be increased by optimization of output coupler transparency, eliminating of Fresnel losses with AR-coatings, as well as, reducing pump radiation losses in the atmosphere with the help of vacuum or nitrogen vapor tubes or shortening of its optical path.

Development of powerful 3-μm fiber laser sources provides the convenient possibility for power scaling of Fe:ZnSe laser technology and wavelength extension using other gain media with 3-μm absorption bands (Fe:CdSe e.g. [26]). Nowadays more than 40 W output power at 3 μm is available with further possibilities of power-scaling [27]. High power, great stability and excellent beam quality of such pump lasers are key factors for the development of Kerr-lens mode-locked systems based, as well, on broadband mid-IR active elements. Widely tunable mid-IR sources are interesting for scientific purposes such as molecular spectroscopy, chemistry and medicine.

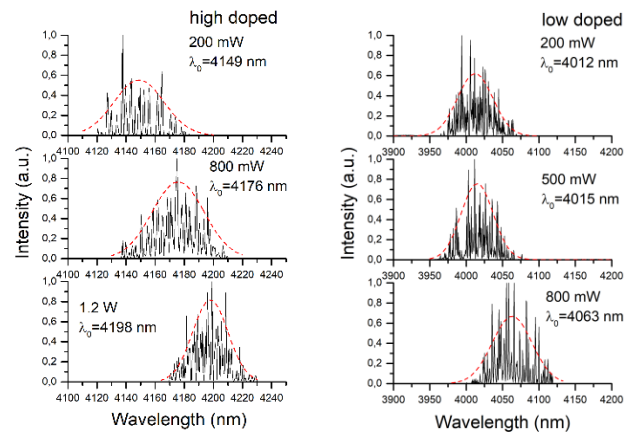

**Fig.4.** Output spectra of Fe:ZnSe laser based on the high (left) and low (right) doped sample for various output power.

**Funding.** Russian Foundation for Basic Research (RFBR) (18-52-50019); JSPS Bilateral Joint Research Project (2018-2020); JSPS Grants-in-Aid for Scientific Research (15K06565);